\documentclass[twoside,twocolumn,9pt]{article}
\usepackage{extsizes}
\usepackage[super,sort&compress,comma]{natbib} 
\usepackage[version=3]{mhchem}
\usepackage[left=1.5cm, right=1.5cm, top=1.785cm, bottom=2.0cm]{geometry}
\usepackage{balance}
\usepackage{mathptmx}
\usepackage{sectsty}
\usepackage{graphicx} 
\usepackage{lastpage}
\usepackage[format=plain,justification=justified,singlelinecheck=false,font={stretch=1.125,small,sf},labelfont=bf,labelsep=space]{caption}
\usepackage{float}
\usepackage{fancyhdr}
\usepackage{fnpos}
\usepackage[english]{babel}
\addto{\captionsenglish}{%
  
}
\usepackage{array}
\usepackage{droidsans}
\usepackage{charter}
\usepackage[T1]{fontenc}
\usepackage[usenames,dvipsnames]{xcolor}
\usepackage{setspace}
\usepackage[compact]{titlesec}
\usepackage{hyperref}

\usepackage{epstopdf}

\definecolor{cream}{RGB}{222,217,201}

\begin{document}

\pagestyle{fancy}
\thispagestyle{plain}
\fancypagestyle{plain}{
\renewcommand{\headrulewidth}{0pt}
}

\makeFNbottom
\makeatletter
\renewcommand\LARGE{\@setfontsize\LARGE{15pt}{17}}
\renewcommand\Large{\@setfontsize\Large{12pt}{14}}
\renewcommand\large{\@setfontsize\large{10pt}{12}}
\renewcommand\footnotesize{\@setfontsize\footnotesize{7pt}{10}}
\makeatother

\renewcommand{\thefootnote}{\fnsymbol{footnote}}
\renewcommand\footnoterule{\vspace*{1pt}%
\color{cream}\hrule width 3.5in height 0.4pt \color{black}\vspace*{5pt}} 
\setcounter{secnumdepth}{5}

\makeatletter 
\renewcommand\@biblabel[1]{#1}            
\renewcommand\@makefntext[1]%
{\noindent\makebox[0pt][r]{\@thefnmark\,}#1}
\makeatother 
\renewcommand{\figurename}{\small{Fig.}~}
\sectionfont{\sffamily\Large}
\subsectionfont{\normalsize}
\subsubsectionfont{\bf}
\setstretch{1.125} 
\setlength{\skip\footins}{0.8cm}
\setlength{\footnotesep}{0.25cm}
\setlength{\jot}{10pt}
\titlespacing*{\section}{0pt}{4pt}{4pt}
\titlespacing*{\subsection}{0pt}{15pt}{1pt}

\fancyfoot{}
\fancyfoot[LO,RE]{\vspace{-7.1pt}\includegraphics[height=9pt]{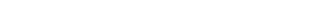}}
\fancyfoot[CO]{\vspace{-7.1pt}\hspace{13.2cm}\includegraphics{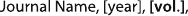}}
\fancyfoot[CE]{\vspace{-7.2pt}\hspace{-14.2cm}\includegraphics{head_foot/RF}}
\fancyfoot[RO]{\footnotesize{\sffamily{1--\pageref{LastPage} ~\textbar  \hspace{2pt}\thepage}}}
\fancyfoot[LE]{\footnotesize{\sffamily{\thepage~\textbar\hspace{3.45cm} 1--\pageref{LastPage}}}}
\fancyhead{}
\renewcommand{\headrulewidth}{0pt} 
\renewcommand{\footrulewidth}{0pt}
\setlength{\arrayrulewidth}{1pt}
\setlength{\columnsep}{6.5mm}
\setlength\bibsep{1pt}

\makeatletter 
\newlength{\figrulesep} 
\setlength{\figrulesep}{0.5\textfloatsep} 

\newcommand{\topfigrule}{\vspace*{-1pt}%
\noindent{\color{cream}\rule[-\figrulesep]{\columnwidth}{1.5pt}} }

\newcommand{\botfigrule}{\vspace*{-2pt}%
\noindent{\color{cream}\rule[\figrulesep]{\columnwidth}{1.5pt}} }

\newcommand{\dblfigrule}{\vspace*{-1pt}%
\noindent{\color{cream}\rule[-\figrulesep]{\textwidth}{1.5pt}} }

\makeatother

\twocolumn[
  \begin{@twocolumnfalse}
{\includegraphics[height=30pt]{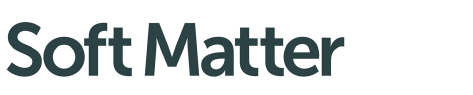}\hfill\raisebox{0pt}[0pt][0pt]{\includegraphics[height=55pt]{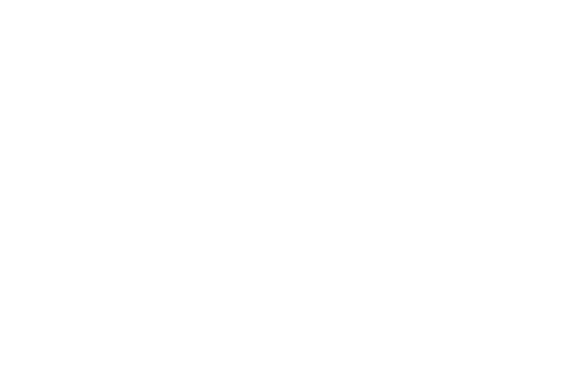}}\\[1ex]
\includegraphics[width=18.5cm]{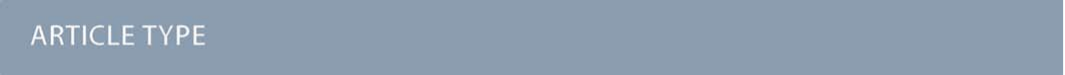}}\par
\vspace{1em}
\sffamily
\begin{tabular}{m{4.5cm} p{13.5cm} }

\includegraphics{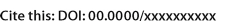} & \noindent\LARGE{\textbf{Dynamics of magnetic self-propelled particles in a harmonic trap$^\dag$}} \\
\vspace{0.3cm} & \vspace{0.3cm} \\

 & \noindent\large{Pamela Mu\~noz Obreque,\textit{$^{a}$} Oscar Garrido,\textit{$^{a}$}  Diego Romero,\textit{$^{a}$}, Hartmut L{\"o}wen,\textit{$^{b}$} and Francisca Guzm\'an-Lastra\textit{$^{a}$}} \\

\includegraphics{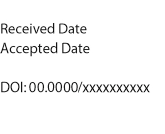} & \noindent\normalsize{Artificial active particles, exemplified by Hexbugs (HB), serve as valuable tools for investigating the intricate dynamics of active matter systems. Leveraging their stochastic motion, which faithfully emulates such systems and their straightforward controllability, Hexbugs provides an excellent experimental model. In this study, we utilize Hexbugs equipped with disk-like armor and embedded magnetic dipoles, transforming them into Magnetic Self-Propelled Particles (MSPP). We explore the emergence of collective and stationary states numerically and experimentally by confining these MSPPs within a parabolic domain acting as a harmonic potential. Our findings unveil a diverse range of metastable configurations intricately linked to the complex dynamics inherent in the system. We discern that particle number, activity, and the balance between magnetic and harmonic potential strengths predominantly influence the metastability of these structures. By employing these parameters as control factors, we compare and contrast the behavior of MSPPs with disk-like magnetic Active Brownian Particles (ABPs) in the overdamped limit of vanishing inertia.
Our numerical predictions reproduce most of the experimental observations,
highlighting the crucial role of magnetic dipole interactions in developing novel
configurations for active particles within parabolic domains. These configurations
include chains, clusters, and vortex formations characterized by a specific pattern in the particle spatial distribution. Notably, we observe that the influence of inertia is not fundamental in generating metastable configurations in these confined systems. Instead, the particle's shape, activity, and orientation are dominant factors. This comparative analysis provides insights into the distinctive features and dynamics of MSPP within confined environments, shedding light on the role of short-range polar interactions.} \\

\end{tabular}

 \end{@twocolumnfalse} \vspace{0.6cm}

  ]

\renewcommand*\rmdefault{bch}\normalfont\upshape
\rmfamily
\section*{}
\vspace{-1cm}


\footnotetext{\textit{$^{a}$~Departamento de F\' isica, Facultad de Ciencias, Universidad de Chile, Santiago Chile. E-mail: fguzman@uchile.cl}}
\footnotetext{\textit{$^{b}$~Institut f{\"u}r Theoretische Physik II: Weiche Materie, Heinrich-Heine-Universit{\"a}t D{\"u}sseldorf, D-40225 D{\"u}sseldorf, Germany. }}

\footnotetext{\dag~Electronic Supplementary Information (ESI) available: [details of any supplementary information available should be included here]. See DOI: 10.1039/cXsm00000x/}

\footnotetext{\ddag~Additional footnotes to the title and authors can be included \textit{e.g.}\ `Present address:' or `These authors contributed equally to this work' as above using the symbols: \ddag, \textsection, and \P. Please place the appropriate symbol next to the author's name and include a \texttt{\textbackslash footnotetext} entry in the the correct place in the list.}



\section{Introduction}
In recent decades, the study of active Brownian particles (ABPs) has captivated considerable attention within the scientific community, thanks to their distinctive non-equilibrium nature. Unlike traditional Brownian particles that rely solely on thermal interactions for movement, ABPs possess the unique ability of self-propulsion, acquiring additional kinetic energy from their surroundings to sustain non-equilibrium states \cite{gompper20202020,romanczuk2012active,bechinger2016active}. This intrinsic property results in intricate and dynamic behaviors, making ABPs a focal point in the field of Active Matter \cite{romanczuk2012active, chaudhuri2014active, pohl2014dynamic, vanesse2023collective, filella2018model}, which endeavors to understand the collective dynamics observed in diverse systems such as bird flocks, fish schools, bacterial colonies \cite{petroff2022phases, cates2012diffusive}, and groups of micro-robots \cite{dauchot2019dynamics, deblais2018boundaries, giomi2013swarming,vutukuri2017rational}.

One remarkable aspect of Active Matter is its adaptability, allowing it to respond to external stimuli such as magnetic fields and magnetic dipole-dipole interactions \cite{dreyfus2005microscopic,vincenti2019magnetotactic,lee2004controlled, shani2014long,liao2021emergent,kogler2015lane, waisbord2016destabilization}, light \cite{desai2017modeling,siebers2023exploiting}, heat \cite{pohl2014dynamic, chaudhuri2014active}, or confinement \cite{araujo2023steering,basu2018active,deblais2018boundaries, faundez2022microbial,ginot2015nonequilibrium,leoni2020surfing}, leading to ever-evolving collective dynamics. The configurations arising from these stimuli often exhibit apparent disorder and transient metastable states, stable for a limited duration and subject to recurrence \cite{wang2023spontaneous, klumpp2019swimming,xie2019reconfigurable,guzman2016fission}. 

Hexbugs (HB), small robots propelled by vibrating motors powered by internal batteries, offer an accessible tool for investigating the dynamics of active systems \cite{Hexbugs, sepulveda2021bioinspired,dauchot2019dynamics,deblais2018boundaries,tapia2021trapped}. While Hexbugs do not extract energy from their environment, they exhibit behaviors akin to artificial active matter systems, showcasing collective motion under specific conditions \cite{sepulveda2021bioinspired,dauchot2019dynamics,deblais2018boundaries,engbring2023nonlinear}. Previous research has explored Hexbugs in various conditions, altering confinement surfaces and introducing potentials like elastic and magnetic interactions \cite{tapia2021trapped,yang2020robust,sepulveda2021bioinspired}. Confining Hexbugs within surfaces such as parabolic or ellipsoidal disks \cite{dauchot2019dynamics,tapia2021trapped, leoni2020surfing,siebers2023exploiting}, these parabolic domains have been often modeled as harmonic potentials, which have shown to regulate dynamics at the edges, preventing particle accumulation at the surface \cite{basu2018active,buttinoni2022active,jahanshahi2017brownian, volpe2013simulation}. 

Here, we explore collective transient and steady configurations under the effect of confinement and magnetic dipole interactions by conducting experimental studies using varying numbers of disk-like Hexbugs within a parabolic domain \cite{dauchot2019dynamics, buttinoni2022active}. Each Hexbug is equipped with a magnet, inducing dipole-dipole interactions. By studying parameters such as spatial distribution, total magnetization, and the radius of gyration, we compare their behavior with that obtained through numerical simulations using disk-like magnetic Active Brownian Particles (ABPs) in the overdamped limit of vanishing inertia \cite{telezki2020simulations}. We reveal a diverse range of metastable configurations intricately linked to system dynamics. Factors such as particle number, activity, and the balance between magnetic and harmonic potentials predominantly influence metastability—the crucial role of magnetic dipole interactions in generating novel configurations within parabolic domains. Chains, clusters, and vortex formations with distinct spatial distributions emerge. 
We observed that although inertia plays a role in particle spatial distribution and typical steady configurations favoring cluster formations for particle number $N>3$,   \cite{deblais2018boundaries,engbring2023nonlinear}, inertia is non-fundamental in generating metastable configurations during particle-particle short interactions \cite{leoni2020surfing}. Instead, particle shape, activity, and orientation play dominant roles. Our numerical predictions partially reproduce experimental observations, providing insights into polar interactions' distinctive features and dynamics within confined environments.

This paper is organized as follows: In Sec.~\ref{sec1}, the basic model for a magnetic active Brownian particle inside a harmonic potential is introduced together with the numerical simulations, and for comparison, a minimum experimental model for magnetic self-propelled particles inside a parabolic domain is described. 
Section~\ref{sec2} contains results and discussions for the average spatial distribution for numerical simulations. Sec.~\ref{sec3} displays transient and steady configurations observed while performing experiments and numerical simulations, and they are discussed in terms of the average gyration radius and total magnetization. Finally, Sec.~\ref{sec4} is devoted to conclusions and outlook.

\section{Magnetic self-propelled particles (MSPP)}
\label{sec1}
\begin{figure}[ht!]
\centering 
\includegraphics[width=\columnwidth]{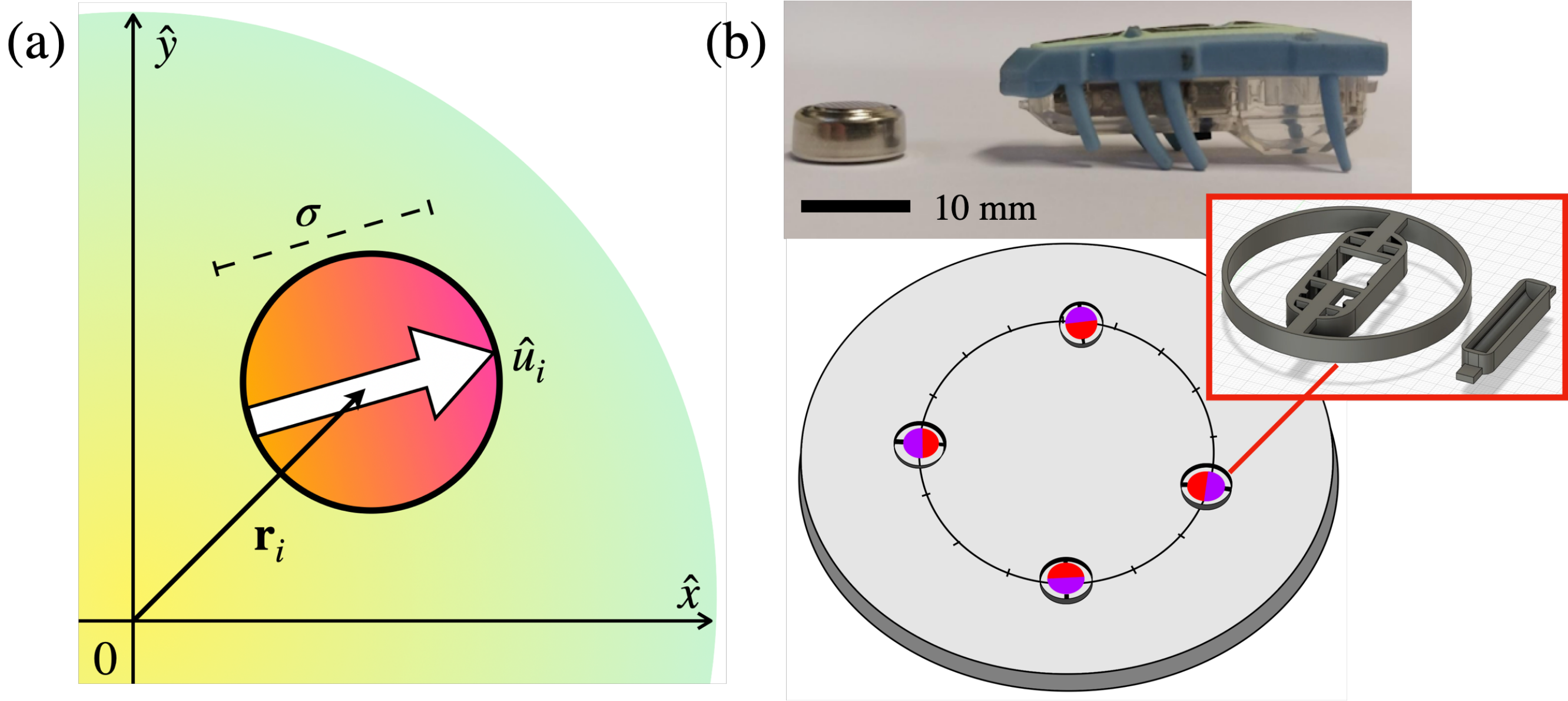}	
    \caption{(a) A Magnetic Self-Propelled Particle (MSPP) shaped like a disk with a diameter of $\sigma$. The MSPP is located on the x-y plane which has its origin at the potential well's center. Each MSPP is characterized by its position $\mathbf{r}_{i}(t)$ and orientation vector $\hat{\mathbf{u}}_i(t)$, which is represented by a white arrow and a color gradient. The orientation vector is defined as $\hat{\mathbf{u}}_i(t)=[\cos\theta_i(t),\sin\theta_i(t)]$. (b) The items shown from top to bottom are: a LR44 battery with a voltage of $1.55$ [V], a Hexbug Nano that is $4.5 \pm 0.1$ [cm] in length, a disk-like armor with a diameter of $D=6.5 \pm 0.1$ [cm] and a cylindrical compartment that is $L_C=4 \pm 0.1$ [cm] in length. Four Hexbug Nano were placed equidistantly from the center of the parabolic dish.} 
\label{fig:esquema}
\end{figure}

We consider a system of $N$ disk-shaped magnetic self-propelled particles (MSPP) that move inside an harmonic potential (HP), denoted as $U^{\text{HP}}$. The position vectors of the particles are represented by $\mathbf{r}_i(t)=[x_i(t),y_i(t)]$, while their orientation vectors are given by $\hat{\mathbf{u}}_i(t)=[\cos\theta_i(t),\sin\theta_i(t)]$ in 2-D. Each particle has a magnetic dipole moment $\mathbf{m}_i=m\hat{\mathbf{u}}_i(t)$ directed along its orientation axis, where $m$ is the magnitude of the point dipole. The overdamped equation of motion for the $i-$th MSPP is,
\begin{eqnarray}
\dot{\mathbf{r}}_i&=&u_{0}\hat{\mathbf{u}}_i(t)-\frac{1}{\gamma_{T}}\nabla_{\mathbf{r}_i}\left (\sum_{j,j\neq i}(U^{\text{WCA}}_{ij}+U^{\text{D}}_{ij})+U^{\text{HP}}_i\right)  \label{eq1}\\
    \dot{\hat{\mathbf{u}}}_i&=&(1/\gamma_R)(\boldsymbol{\xi}_{i,R}(t)-\mathbf{T}_i)\times\hat{\mathbf{u}}_i\label{eq2}
\end{eqnarray}

\noindent where $\gamma_T$ and $\gamma_R$ are the friction coefficients associated with translation (T) and rotation (R) in the $x-y$ plane, respectively. Here, $u_{0}$ represents the self-propelled velocity. 
$\boldsymbol{\xi}_{i,R}(t)$ represents the rotational Gaussian white noises of zero mean, with $\langle\boldsymbol{\xi}_{j,R}(t)\rangle=0$. The time correlation is $\langle\boldsymbol{\xi}_{j,R}(t_1)\boldsymbol{\xi}_{j,R}(t_2)\rangle=2 k_BT/\gamma_R\delta(t_1-t_2)$. We define $D_R$ as the rotation diffusion parameter, which is $D_R=\frac{2k_BT}{\gamma_R}$, where $k_BT$ denotes an effective thermal energy. Additionally, each MSPP is subject to three types of potentials that in turns, represents three forces: an excluded volume potential ($U^{\text{WCA}}$) modeled with the Weeks-Chandler-Anderson (WCA) potential, a magnetic dipole-dipole interaction ($U^{\text{D}}$), and an harmonic potential (HP), ($U^{\text{HP}}$). Where, $\mathbf T_j=\hat{\mathbf{u}}_j\times\nabla_{\hat{\mathbf{u}}}U^{\text{D}}$ is the magnetic torque acting on particle $j$ and the potential energies are given,

\begin{equation}\label{eq.4}
 U^{\text{WCA}}_{ij} =  \left\{\begin{array}{cc}
     \displaystyle{4 \varepsilon\left[\left(\frac{\sigma}{r_{i j}}\right)^{12}-\left(\frac{\sigma}{r_{ij}}\right)^{6}\right]}    &   r_{i j} \leq r_{m}\\
     0    & \mbox{otherwise} 
    \end{array} \right.
\end{equation}

\begin{equation}
    U^{\text{D}}_{ij}=\frac{m^2}{r_{ij}^3}\left[\hat{\mathbf{u}}_i\cdot \hat{\mathbf{u}}_j-\frac{3(\hat{\mathbf{u}}_i\cdot \mathbf r_{ij})(\hat{\mathbf{u}}_j\cdot\mathbf{r}_{ij})}{r_{ij}^2} \right]
\end{equation}

\begin{equation}\label{HP}
    U^{\text{HP}}_i=\frac{\kappa}{2}r_i^2
\end{equation}
    
\noindent where $\sigma$ corresponds to the particle diameter, $\mathbf r_{ij}=\mathbf r_j-\mathbf r_i$ is the position of particle $j$ relative to particle $i$ and $r_{ij}$ their distance , $r_m=2^{-1/6}\sigma$ is the potential cut-off interaction, $\varepsilon$ is an energy interaction strength.
Finally, in Equation~\eqref{HP} $U^{\text{HP}}$ represents the harmonic potential (HP) used to model an optical tweezers \cite{buttinoni2022active} or a parabolic surface where particles are bounded \cite{dauchot2019dynamics, engbring2023nonlinear}, with strength $\kappa$, located at the origin of the coordinate system, see Figure~\ref{fig:esquema}(a).

\subsection{Numerical simulations}
We use standard Brownian dynamics methods to numerically solve the governing Eqs.\eqref{eq1} and \eqref{eq2} for $N=2,3,4$ and $5$ MSPP. The particle diameter $\sigma=2R$ is use as the unit of length, $\tau=\gamma_T \sigma^2/4\varepsilon$ as the unit of time and $\varepsilon$ is the energy scale.
We fixed the friction coefficients $\gamma_T=6\pi$ and $\gamma_R=8\pi$. We integrated the equations of motion Eqs. \eqref{eq1} and \eqref{eq2} with a time step of $\Delta t=10^{-4}\tau$, using the third-order Adams-Bashforth-Moulton predictor-corrector method for a simulation time of $T_s=10^4\tau$. In these units, the system behavior is characterized by four different dimensionless parameters: the reduced magnetic dipole strength $m^*=m/\sqrt{\varepsilon \sigma^3}$, the reduced harmonic potential strength $\kappa^*=\kappa/\varepsilon \sigma^2$, the reduced rotational diffusivity $D_R^*=D_R\tau$, and the reduced self-propelled velocity $u_0^*=u_0\tau/\sigma$. We fix the rotational diffusivity to $D_R^*=10^{-2}$ and set the reduced $\kappa^*$ to two different strengths, namely $\kappa^*=(10^{-2}, 10^{-3})$, which define a strong and a weak trap. We varied the reduced magnetic moment $m^*$ in the range $(0.7-2)$ and the self-propelled velocity $u_0^*$ in the range $(1-10)$.
\subsection{Experiments}
We experimentally modeled a single MSPP using a Hexbug Nano \cite{Hexbugs} with a disk-like armor printed in 3D with diameter $D=6.5 \pm 0.1$ [cm] and mass $M_{A}=5 \pm 0.1$ [g]. Along the diameter of the armor, we put a cylindrical compartment with length $L_{C}=4 \pm 0.1$ [cm] and mass $M_{C}=1 \pm 0.1$ [g]. See Figure~\ref{fig:esquema}(b). We implanted a Neodymium rod magnet with diameter $D_{N}=6 \pm 0.01$ [mm] and length $L=3.5 \pm 0.1$ [cm] on the compartment of each disk. 
We numerically varied different magnetic moments $m$ close to the prior reported magnetic moment for a Neodymium rod magnet in the literature, which is approximately $m_{exp}=0.04$ A m$^{^2}$ \cite{sepulveda2021bioinspired}. In order to compare our numerical simulations with the experimental realizations, we will consider the reported potential interaction strength between two hexbugs, which varies between $\varepsilon_{exp}=5-10$ Nm \cite{sepulveda2021bioinspired}, then a good approximation to the adimensional magnetic moment is $m^*_{exp}=m_{exp}/\sqrt{\varepsilon_{exp}D^3}$ which give us variations between $m^*_{exp}\approx 0.7-1$.

We tested each MSP-Hexbug by letting it move on a flat surface to ensure it performs a persistent Brownian motion instead of a chiral-like one. We assumed that all magnetic self-propelled Hexbugs (MSP-Hexbugs) move on average with the same constant velocity $u_0$. During experiments, we observed that MSP-Hexbugs varied their velocity due to battery drain, which occurs over some time longer than the experiment's time performance. Therefore, we used unused batteries of $1.54 \pm 0.01$ [V] for all experiment realizations to keep a constant average velocity $u_0\approx5$[cm/s] \cite{sanchez2018self}, which gives us an estimation for the adimensional velocity during experiments close to $u_{0,exp}^*\approx 2.6$ if we consider a characteristic time scale during experiments close to $\tau_{exp}\approx 3.3$ s as was reported in \cite{tapia2021trapped}.

All MSPPs were free to move inside a satellite dish with a parabolic domain, with dimensions of $L_x=70$, $L_y=65$, and $L_z=5.5$ [cm]; the restitution coefficient for the parabolic dish is $\kappa_{exp}\approx M_A g L_z/L_xL_y=0.01$[N/m], then $\kappa_{exp}^*=\kappa_{exp}/\epsilon_{exp}D^2\approx0.28$. During experiments, we released all the MSPPs radially at a fixed distance of $5$ [cm] from the parabolic center (see Figure~\ref{fig:esquema}(b)). We recorded their motion using a standard cellphone camera, and after the experiment, we analyzed all the videos, programming a code in Python to extract their trajectories and orientations.

\begin{figure}[ht!]
\centering
\includegraphics[width=\columnwidth]{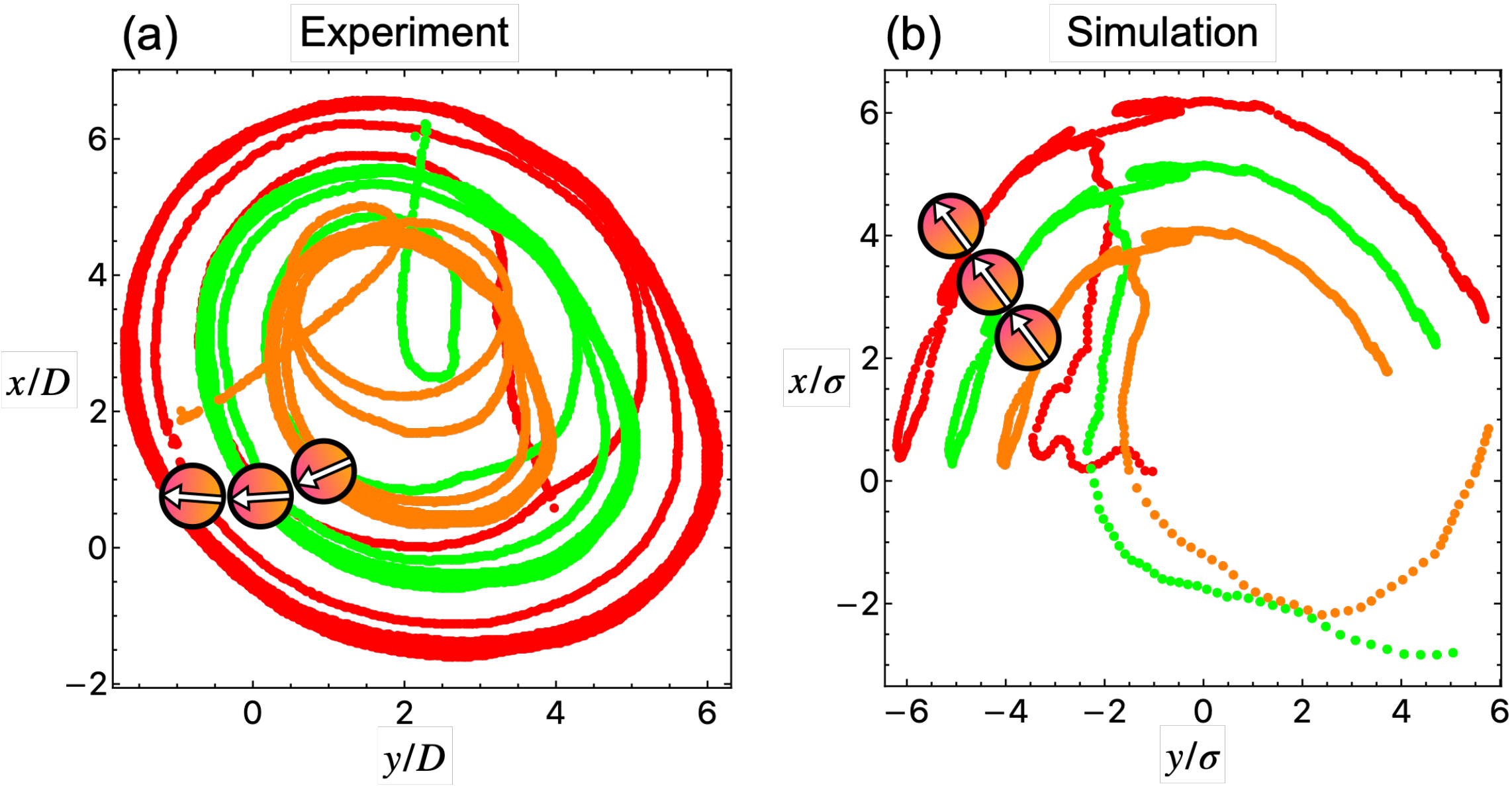}
\caption{(a) Experimental trajectory for $N=3$ MSP Hexbugs. Hexbugs were initially placed radially toward the parabolic dish center. They form a chain-like configuration that points radially towards to the parabolic domain frontier. The red, green, and orange lines correspond to each particle's trajectories. (b) Simulated trajectory for $N=3$ MSPP. Particles were initially positioned at random positions inside the harmonic potential. A particle chain is observed pointing radially from the harmonic potential origin and moving tangentially with respect to their orientation.}
\label{fig:tray}
\end{figure}

\section{Average spatial distributions inside a parabolic domain}
\label{sec2}
We conducted numerical simulations to study the effects of varying the trap strength $\kappa^*$, magnetic moment $m^*$, and self-propelled velocity $u_0^*$ while keeping the rotational diffusivity $D_R^*$ constant for all MSPP. We then studied the time-averaged spatial distribution for $N$ particles inside the harmonic potential by analyzing their trajectories.

Figure~\ref{fig:tray}(a) shows the rescaled trajectories for experiments with  $N=3$ MSP-Hexbugs forming a chain-like structure pointing radially outside the parabolic domain. The particles explore the domain by moving perpendicular to the frontier of the parabolic dish, reproducing the first steady state reported in \cite{dauchot2019dynamics}.

Figure~\ref{fig:tray}(b) depicts a similar scenario for simulations with  $N=3$ MSPP. In the case of a weak trap, with $m^*=2$ and self-propelled velocity $u_0^*=3$. The particles again explore the confinement by moving radially from the harmonic trap's origin and tangentially with respect to their orientation.

\begin{figure}[ht!]
    \centering 
\includegraphics[width=1\columnwidth]{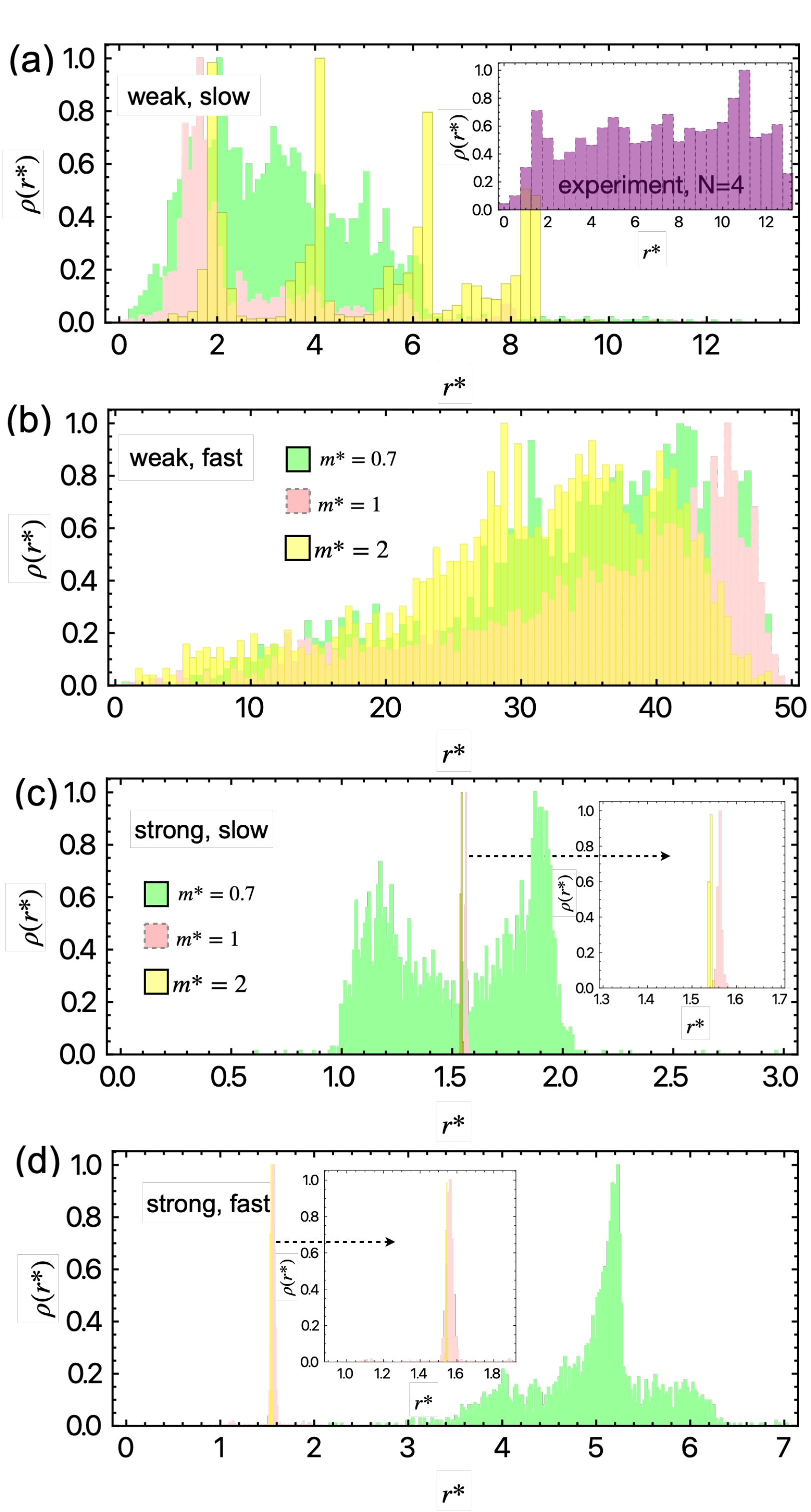}	
    \caption{Average spatial distribution for the numerical simulations, denoted as $\rho(r^*)$ with $r^*=r/R$, for a $N=4$ Magnetic Self-Propelled Particles (MSPP) system. In the plots, green bars represent the results for numerical simulations with weak magnetic dipole-dipole interaction ($m^*=0.7$) and pink bars in case $m^*=1$. In contrast, yellow bars depict the results for numerical simulations with strong dipole-dipole interaction case ($m^*=2$). The inset in panel (b) is the average spatial distribution (in purple bars) for an experiment with $N=4$ MSP-hexbugs.
The different panels illustrate distinct scenarios:
(a) Average spatial distribution in the case of a weak harmonic trap ($\kappa^*=10^{-3}$) and slow MSPP ($u_0^*=1$).
(b) Average spatial distribution in the case of a weak harmonic trap ($\kappa^*=10^{-3}$) and fast MSPP ($u_0^*=10$).
(c) Average spatial distribution in the case of a strong harmonic trap ($\kappa^*=10^{-2}$) and slow MSPP ($u_0^*=1$).
(d) Average spatial distribution in the case of a strong harmonic trap ($\kappa^*=10^{-2}$) and fast MSPP ($u_0^*=10$).} 
\label{fig:pdf}
\end{figure}

In Figure~\ref{fig:pdf}, we present the average spatial distribution given by
\begin{equation}
\rho(r)=\langle\overline{\sum_{i=1}^N\delta(r_i-r)}\rangle,
\end{equation}
where $\langle\cdot\rangle$ represents a noise average and $\overline{\cdots}$ is a time average for the case of $N=4$ MSPP, extracted from numerical simulations. We generated a histogram for particle positions $r$ measured with respect to the harmonic potential center (see Figure~\ref{fig:esquema}). The particle spatial distributions depend significantly on the trap strength $\kappa^*$ and the magnetic dipolar moment $m^*$ relative to the self-propelled velocity $u_0^*$.

We use green bars to represent weak magnetic dipole interactions with $m^*=0.7$, pink bars for $m^*=1$, and yellow bars for the strongest magnetic dipole moment, $m^*=2$. While slow particles tend to accumulate at the center of the optical trap, displaying layering in different shells, the number and sharpness of peaks depend on $m^*$.
For the case of a weak harmonic trap and slow particles (Figure~\ref{fig:pdf}(a)), which corresponds to values close to our experimental parameters, there are four shells with strongly asymmetric radial peaks, consistent with particles moving radially outward until they reach a maximum distance from the optical trap's center. The radial peaks become smeared out with smaller magnetic interactions (green data in Figure~\ref{fig:pdf}(a)(c)(d)), whereas a strong optical trap and large magnetic dipole interaction result in a sharper single peak with all particles at the same distance from the center of the optical trap (see yellow and pink bars in Figure~\ref{fig:pdf}(c)(d)).
None of the distributions are Gaussian; instead, we observe a standard peak-like distribution inside the optical trap. Our experiments show that MSP-Hexbugs tend to explore the parabolic dish spatially, spending long times pointing radially outward (see inset in Figure~\ref{fig:pdf}(b) in purple bars for $N=4$ MSP-hexbugs) \cite{dauchot2019dynamics, giomi2013swarming}.
This behavior does not impede the particles from engaging in clustering behaviors due to their magnetic dipole interactions, forming stable configurations, and collectively moving within the confinement of the satellite dish. To visually capture and illustrate this behavior, we have included Supplementary Movie 1.
Both self-propelled velocity and harmonic potential strength affect particle displacement relative to the trap's center. At the same time, magnetic dipole-dipole interactions play a crucial role in defining the system's final state. We will discuss these additional insights in the next section.

\section{Transient and steady states}
\label{sec3}
During our experiments and simulations, we observed a variety of stable configurations along with transient configurations where particles formed interesting patterns. These patterns included chains, vortex, and clustered arrangements. These short-lived configurations were characterized by particles clustering together for brief periods before disbanding into transiently stable pairs.

\begin{figure}[ht!]
    \centering
    \includegraphics[width=\columnwidth]{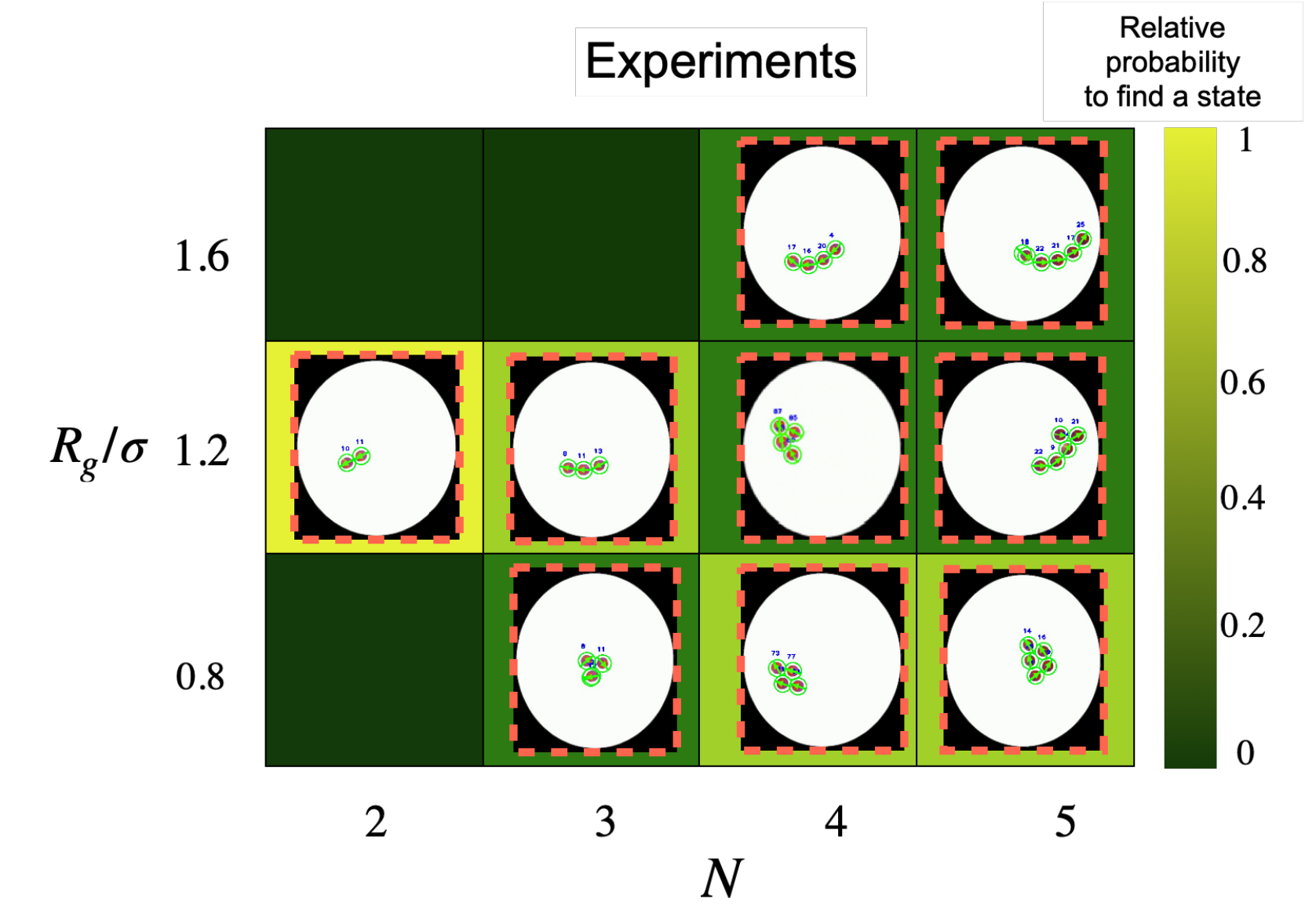}
    \caption{Characterization of the normalized recurrence event in different metastable configurations for experiments with $N=2,...,5$ MSP-Hexbugs. The gyration radius relative to their diameter, denoted as $R_g/\sigma$, was used for this characterization. The observed states are represented from light green to dark green as a relative probability to find a state.
    In the case of $N=2$, a chain-like configuration was observed, as depicted in that specific row. Moving on to $N=3$, two distinct metastable states emerged—a vortex configuration and a chain configuration. For $N=4$, three metastable configurations were observed, including two cluster-like configurations and a bent chain. Finally, in the case of $N=5$, three configurations were identified a cluster-like formation, a Y-junction \cite{kaiser2015active}, and a bent chain.}
    \label{fig:mapadecalor}
\end{figure}

Regarding experiments, we conducted meticulous observations, taking into account the time aspect. As particles remained for longer periods, distinct configurations emerged. We attempted to unravel the underlying patterns by characterizing these configurations based on the number of particles, denoted as $N$, and the recurrence event for different configurations observed during experiments represented by the gyration radius, here representing a compartment parameter, see Figure~\ref{fig:mapadecalor}. Particularly, in the case of $N=4$ MSP-hexbugs, we observed a cluster formation with higher recurrence event (see Supplementary movie 1), which displays a spatial distribution with r-spaced peaks as we observed in Figure~\ref{fig:pdf}(b), that resembles in simulations the case of slow MSPP inside a weak harmonic trap and strong magnetic dipole interactions $m^*=2$ (see Figure~\ref{fig:pdf}(a)).
It is noteworthy that even with constant magnetic dipole-dipole interaction, a fixed parabolic domain, and uniform self-propelled velocity, the dynamics of particle encounters were intricately linked to their initial conditions (see Supplementary Movie 2). Therefore, we set up our experiments with particles starting their movement from a fixed radial distance relative to the center of the satellite dish and maintaining a uniform separation distance between the MSP-Hexbugs. This meticulous setup allowed us to delve into the nuanced interplay of forces and initial conditions governing the captivating dynamics within our experimental arena.

\subsection{Gyration radius}
One effective method to study dynamic and steady states observed in both experiments and simulations is through the radius of gyration, denoted as $R_g$. This parameter is defined as the distance of particle $i$ relative to the center of mass of the system composed of $N$ particles, expressed as:
\begin{equation}
    R_g^2=\frac{1}{N}\langle\sum_{i=1}^N{(r_i-r_{cm})^2}\rangle
\end{equation}
Here, $r_{cm}=\frac{1}{N}\langle\sum_{j=1}^N{r_j}\rangle$ represents the center of mass of the system, with a fictive unit mass attributed to each MSPP. 
\begin{figure}[ht!]
    \centering
    \includegraphics[width=\columnwidth]{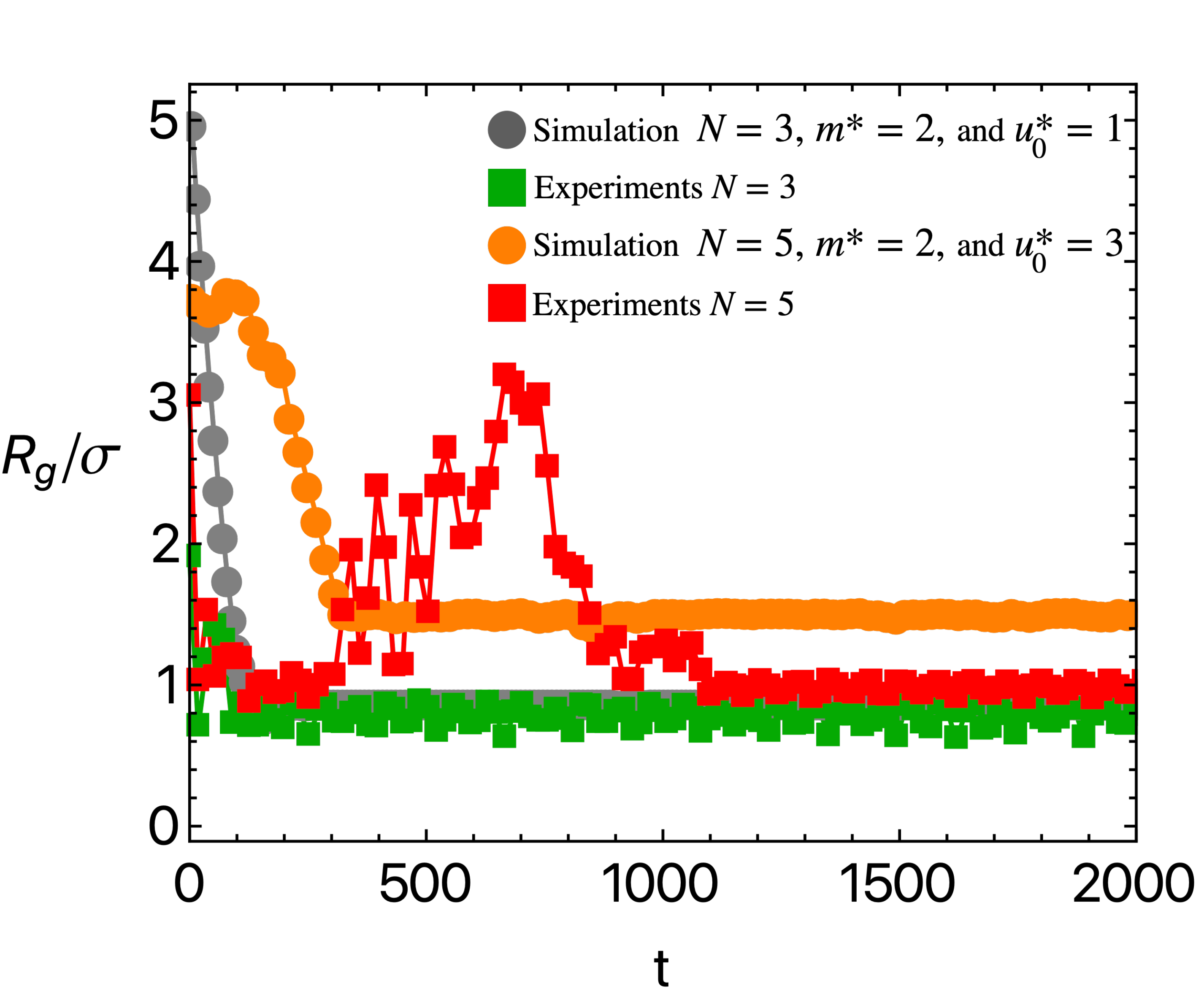}
    \caption{The time evolution of the gyration radius is illustrated for experiments with $N=3$ MSP-Hexbug (green squares) and $N=5$ MSP-Hexbug (red squares). Corresponding simulation results for the same particle numbers are presented with $N=3$ MSPP (gray disks) and $N=5$ MSPP (orange disk). For $N=3$ particles, both in experiments and simulations, the system converges to a chain configuration. However, for $N=5$, while simulations show the system forming a chain, experiments reveal the emergence of a cluster-like formation, as depicted in Figure~\ref{fig:mapadecalor}.}
    \label{fig:rg}
\end{figure}
Experimental and numerical validation demonstrates that $R_g$ values close to a fraction of $\sim N\sigma$ indicate interaction among all MSPP, while large values for $R_g$ imply fragmented dynamics.
In Figure~\ref{fig:rg}, we depict the time evolution of $R_g$ divided by the particle diameter $\sigma$ during simulations. For $N=3$ particles (yellow disk) and $N=5$ particles (orange disk) with $\sigma=2$, green and red squares represent experimental results for $N=3$ and $N=5$ MSP-hexbugs, respectively. Here, $\sigma=D=6.5$ [cm] for MSP-hexbugs. The gyration radius exhibits short-time fluctuations at the beginning of realizations, where particles form various transient configurations with small $R_g$ and then, at intermediate times, display fragmented dynamics with large $R_g$ values (refer to Supplementary Movie 3). Subsequently, all realizations converge to a constant $R_g$ value, signifying the attainment of a steady state.
For $N=5$, simulations show the system reaching a chain formation, whereas experiments reveal a cluster-like formation (see Figure~\ref{fig:mapadecalor}). Conversely, for $N=3$, both simulations and experiments converge to a chain-like formation with a constant $R_g$ value.

\subsection{Magnetization}
As an additional order parameter for characterizing both steady and dynamic configurations, we investigated the total magnetization or polarization, denoted as $M$, within the system. It is defined as:

\begin{equation}
    M = \frac{1}{N}\left|\langle\sum_{i=1}^{N}\hat{\mathbf{u}}_i(t)\rangle\right|
\end{equation}

Here, $\hat{\mathbf{u}}_i(t)=[\cos\theta(t),\sin\theta(t)]$ represents the orientation vector for each MSPP, and $N$ is the total number of particles. A magnetic moment equal to $M=1$ indicates a configuration where all particles align in the same direction, resembling a chain-like or cluster-like arrangement, regardless of the particle number $N$. On the other hand, $M=0$ characterizes a vortex-like configuration or a split cluster, where particles point radially outside the parabolic domain (refer to Figure~\ref{fig:mapadecalor}). Intermediate values could represent either stable or transient configurations.

\begin{figure}[ht!]
    \centering
    \includegraphics[width=\columnwidth]{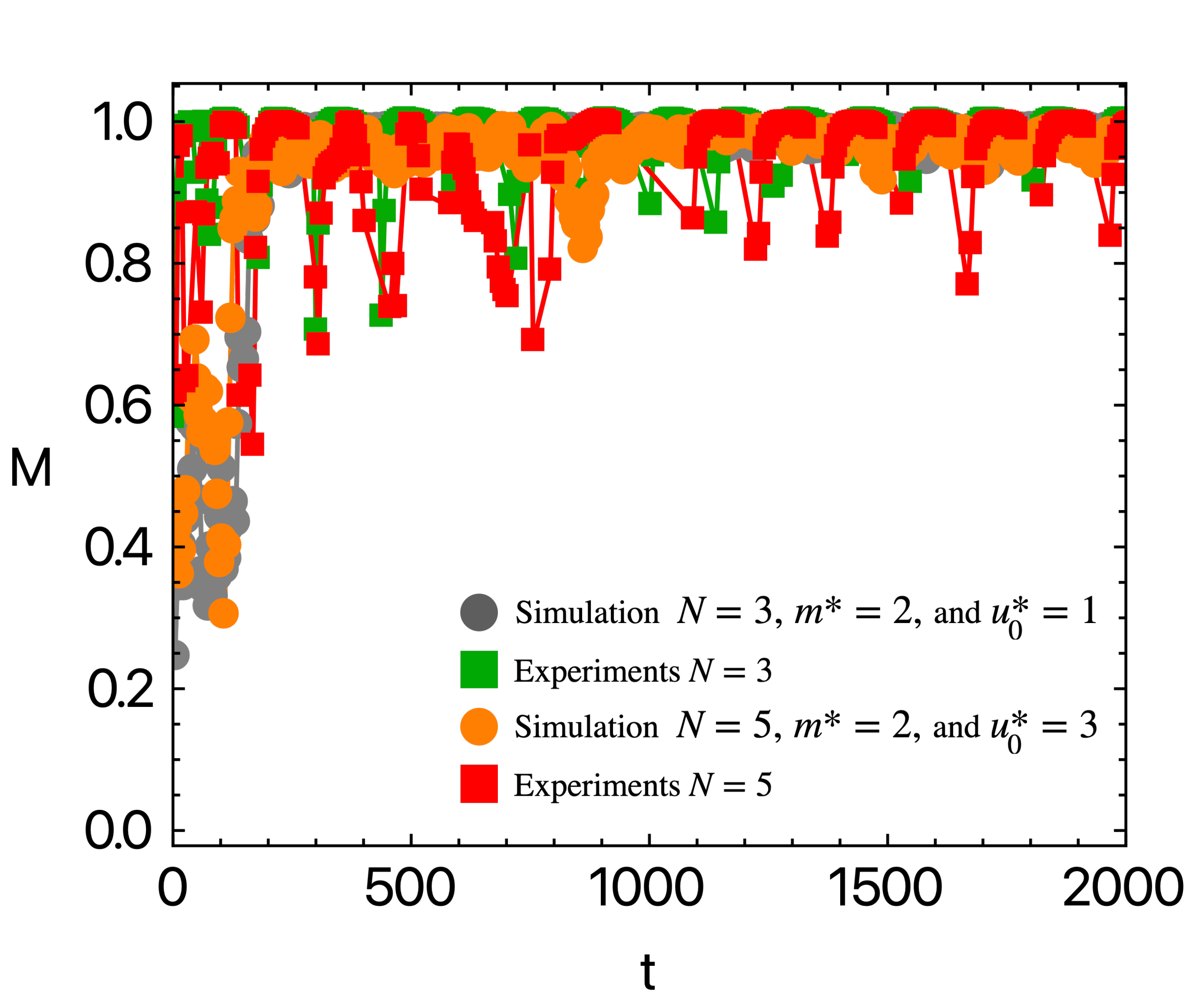}
    \caption{The time evolution of the total magnetization is presented for experiments involving $N=3$ MSP-Hexbugs (green squares) and $N=5$ MSP-Hexbugs (red squares). Corresponding simulation results for the same particle numbers are presented with $N=3$ MSPP (gray disks) and $N=5$ MSPP (orange disks). A transient behavior is observed at short times with all the presented cases converging to a constant value equal to $M\approx1$ corresponding to a chain formation or a cluster with all particles pointing in the same direction.}
    \label{fig:mag}
\end{figure}
In Figure~\ref{fig:mag}, we present the time evolution of the total magnetization $M$ for the cases discussed in the previous section (see Figure~\ref{fig:rg}). Notably, significant fluctuations are observed at the onset of each realization, corresponding to transient configurations. However, over an extended period, all studied cases converge to a steady state with small fluctuations around $M=1$.
Notice, that the magnetic moment by itself is not a parameter that can determine the system configuration. Nevertheless, together with the gyration radius is possible to characterise the dynamic and final system configurations \cite{guzman2016fission, liao2021emergent}. 

\begin{figure}[ht!]
    \centering 
\includegraphics[width=\columnwidth]{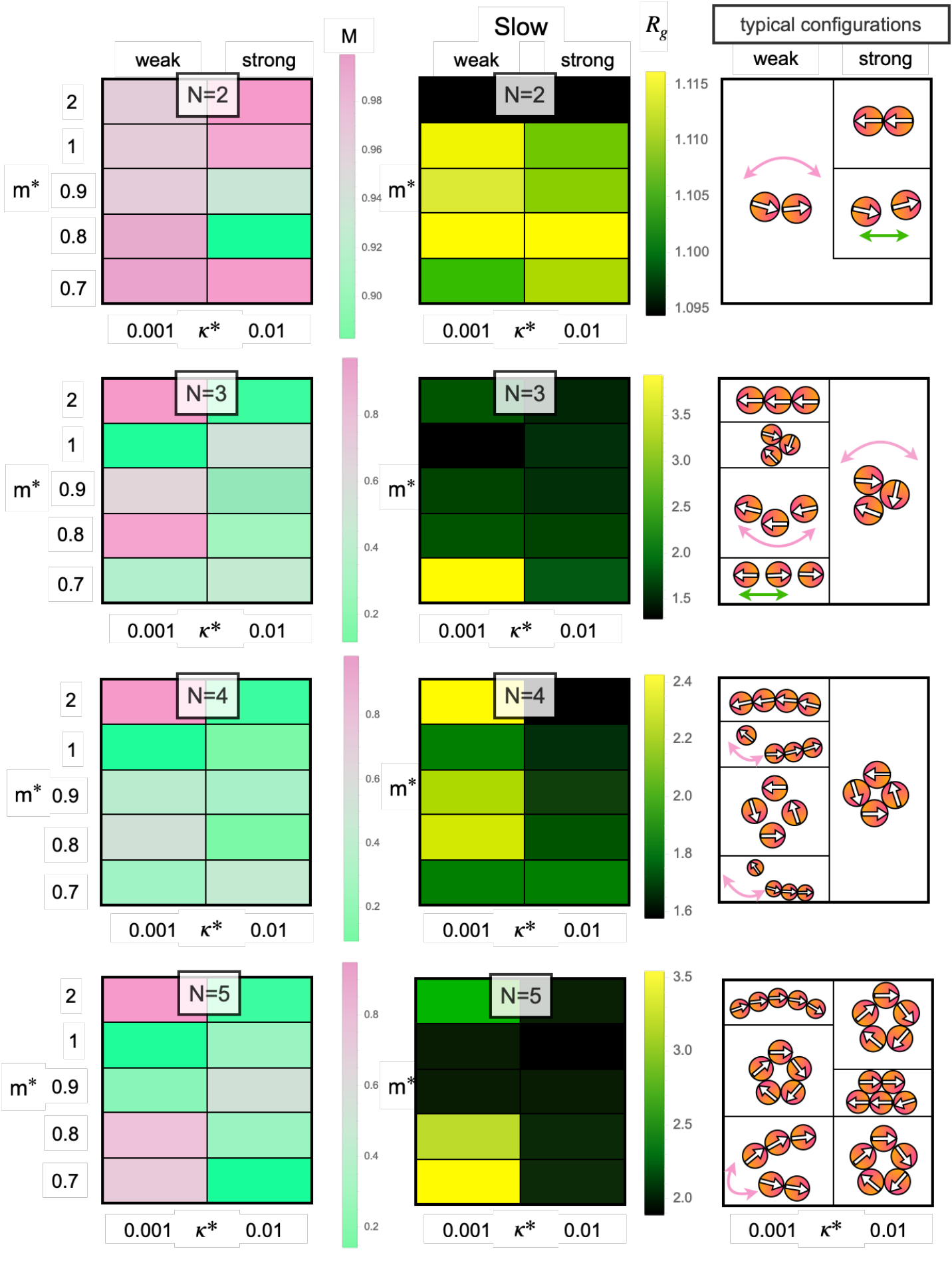}	
    \caption{Phase diagram based on simulation data is presented for a fixed self-propelled velocity ($u_0^*=1$) as a function of MSPP number ($N$) and magnetic dipole interaction ($m^*$). The diagram is constructed from density plots based on the total magnetization ($M$), gyration radius ($R_g$) and typical states for both a strong harmonic potential ($\kappa^*=10^{-1}$) and a weak harmonic potential ($\kappa^*=10^{-2}$). The left column illustrates the total magnetization $M$, while the middle column showcases the corresponding and complementary gyration radius; finally the right column displays the observed steady configurations. The pink arrows represent orientation changes and green arrows distance changes between particles.}
\label{fig:pd1}%
\end{figure}

\begin{figure}[ht!]
    \centering 
\includegraphics[width=\columnwidth]{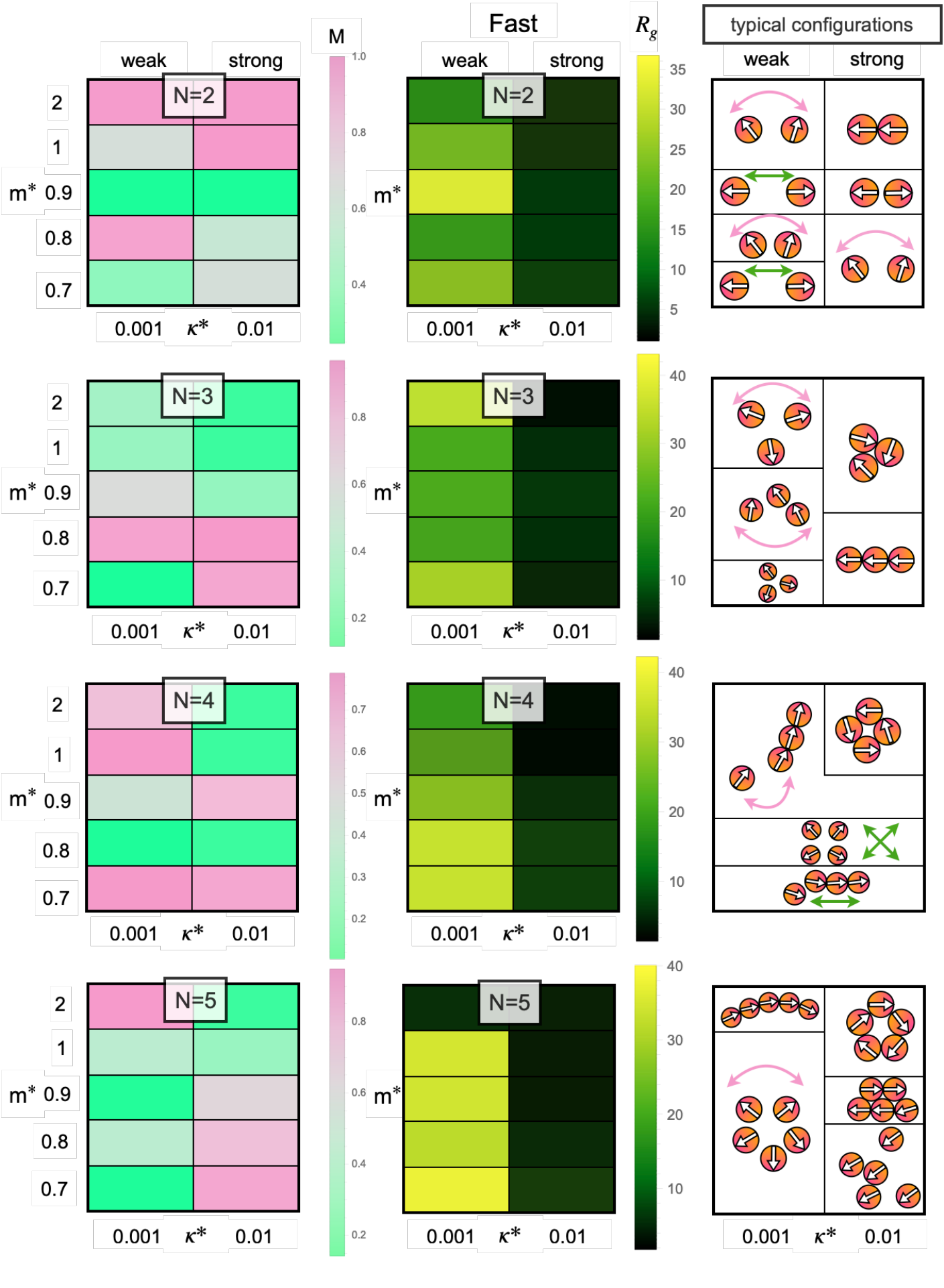}	
    \caption{Phase diagram based on simulation data for a fixed self-propelled velocity ($u_0^*=10$) as a function of MSPP number ($N$) and magnetic dipole interaction strength ($m^*$). The diagram includes information on total magnetization ($M$), gyration radius ($R_g$), and typical configurations for both a strong harmonic potential ($\kappa^*=10^{-1}$) and a weak harmonic potential ($\kappa^*=10^{-2}$). The left column illustrates the total magnetization $M$, the middle column, the complementary gyration radii, and in the right column, the steady configurations are displayed. The pink arrows represent orientation changes and green arrows distance changes between particles.}
\label{fig:pd2}%
\end{figure}

We qualitatively studied these final configurations based on the total magnetic moment and the gyration radius of the final configuration. To systematically analyze the outcomes, we constructed a phase diagram encompassing all particle numbers $N$ and magnetic moments $m^*$ using numerical simulation results for the gyration radius $R_g$ and total magnetization $M$. The density plots are presented in Figures~\ref{fig:pd1} and \ref{fig:pd2}, where each column represents results for strong ($\kappa^*=10^{-2}$) and weak ($\kappa^*=10^{-3}$) optical traps, respectively. Figure~\ref{fig:pd1} corresponds to slow particles with $u_0^*=1$, and Figure~\ref{fig:pd2} represents the case of fast particles with $u_0^*=10$.
In Figure~\ref{fig:pd1} for slow particles, strong confinement leads to the formation of vortex and cluster-like configurations, reminiscent of the behavior observed in active Brownian particles under strong confinement \cite{takatori2016acoustic, buttinoni2022active, dauchot2019dynamics}. Brownian particles tend to accumulate near the optical trap center, as seen in Figure~\ref{fig:pdf}. Interestingly, ring-like vortex configurations also emerge, characterized by particles rotating around the origin with a total magnetization close to 0 and a small but finite gyration radius. Remarkably, these states align with experimental observations, as is presented in Figure~\ref{fig:mapadecalor}. For the case of $N=2$ MSPP with a large magnetic moment $m^*=2$, a stable chain-like configuration is observed, distinct from a ring-like dimer \cite{telezki2020simulations}.

We observe that under weak confinement, chain-like structures emerge as the predominant configurations for all particle numbers and various magnitudes of magnetic dipole interactions. These configurations are characterized by total magnetic moments close to one, as depicted in the pink shadow in Figure~\ref{fig:pd1}, and finite gyration radii larger than 1, presented in Figure~\ref{fig:pd1} as light green regions. The emergence of chain configurations is a novel phenomenon in the context of optical traps with active particles, with r-spaced spatial distributions for each particle, as it was observed in the case of $N=4$ MSPP in Figure~\ref{fig:pdf}(a) in yellow bars. We posit that these structures are primarily formed due to the interplay of magnetic dipole interactions and the inherent tendency of active particles to reorient, pointing radially outward from the harmonic potential \cite{deblais2018boundaries, matas2010hydrodynamic, telezki2020simulations,giomi2013swarming}.

Finally, we considered the case of fast particles with a constant velocity of $u_0^*=10$ under both strong and weak confinement for various magnetic moment strengths ($m^*$) and particle numbers ($N$). In Figure~\ref{fig:pd2}, we presented the total magnetization ($M$) and gyration radius ($R_g$) for the observed final configurations.
Notably, under strong confinement ($\kappa^*=10^{-2}$), when there are two MSPPs, chain configurations are observed. For $N=3$ and weak magnetic interaction, a cluster formation is prominent (pink region), while for a large magnetic moment ($m^*\approx 2$), a ring-like vortex configuration is observed. The case of $N=4$ forms vortex and cluster-like structures (see Figure~\ref{fig:pdf}(d)). For $N=5$ and weak magnetic dipole interactions, split chain configurations or clusters with particles pointing in the same direction are observed, as indicated by the pink region in the total magnetization ($M$).
Under weak confinement, there is a noticeable tendency towards particles remaining at far distances from each other, characterized by light green areas in the gyration radius. The total magnetic moment ($M$) exhibits specific features in these scattered configurations. In many cases, $M=0$ corresponds to all particles pointing radially outward from the optical trap and exploring a spatial region farther from the optical trap's center, as seen in Figure~\ref{fig:pdf}(b). Figure~\ref{fig:pd2} right column displays typical configurations observed during simulations. The pink arrows represent how particles' orientations change between configurations and during time, and the green arrows represent the distance between particles; particles do not conform to steady configurations such as clusters, chains, and vortices, but their configuration changes steadily as a function of time.

\section{Conclusions}
\label{sec4}
In this paper, we explored the characteristics of magnetic active matter on the particle-resolved scale. In detail, we performed numerical simulations to investigate the intricate interplay among confinement, characterized by a harmonic potential, magnetic dipole interactions, and activity. This investigation aimed to uncover the nuanced aspects of spatial distribution, transient dynamics, and steady configurations exhibited by Magnetic Self-Propelled Particles (MSPP) constrained to move inside an optical trap. 
Our findings unveiled a striking parallelism between the analytical framework of overdamped dynamics and experimental results obtained from low-cost experiments involving hexbug particles. These artificial particles, equipped with disk-shaped armor embedded with a magnetic dipole, were confined to a parabolic domain. The synergy of confinement, magnetic interactions, and self-propelled motion showcased intriguing emergent phenomena, underlining the complexity and richness of the observed dynamics.

The average spatial distribution $\rho(r)$ for MSPP in simulations revealed a strong dependence on trap strength ($\kappa^*$), magnetic dipole moment ($m^*$), and self-propelled velocity ($u_0^*$). Regardless of whether $\kappa^*$ was strong or weak and MSPP were slow, particle spatial distributions remained confined to an area close to the harmonic potential center. Notably, even when particles accumulated near the trap center, deviations from expected Gaussian distributions were observed, attributed to the interplay between velocity and magnetic effects.

During simulations under weak confinement and in experiments involving particle numbers $N\leq 3$, chain-like structures emerged as predominant, suggesting a novel phenomenon influenced by magnetic dipole interactions and the inherent reorientation of active particles. Conversely, strong confinement resulted in a vortex and cluster-like configurations. Similar structures were observed experimentally, with clusters pointing radially out of the parabolic dish and chain structures exhibiting higher probability.

Exploring the total magnetization ($M$) provided insights into the system's dynamics, revealing fluctuations at the beginning of realizations before reaching a steady state. The combination of $R_g$ and $M$ enabled the construction of a phase diagram, offering a comprehensive view of system configurations for different particle numbers and magnetic moments.

Although MSP-Hexbugs are considered inertial \cite{tapia2021trapped} and exhibit spatial distributions and trajectories well-described in literature \cite{gutierrez2020inertial, leoni2020surfing}, our minimal model without inertia for slow particles accurately reproduced some observed transient and steady configurations. However, it becomes apparent that steady configurations in both simulations under the overdamped limit and experiments with magnetic hexbugs emerge as a delicate relation between magnetic dipole interactions and self-propulsion.

Our findings highlight the critical interplay among physical parameters such as trap strength, magnetic moment, and self-propelled velocity. These offer valuable insights into the emergent states of Magnetic Self-Propelled Particles (MSPP). Experimental validation with MSP-Hexbugs reveals distinct tendencies for spatial distributions and stable configurations, showcasing the versatility and complexity inherent in active matter systems.

\section*{Author Contributions}
 All authors contributed equally analyzing the data, and writing the manuscript; F.G.-L and H.L. propose the research, F.G.-L. performed the simulations. P.M.O, O.G and D.R performed the experiments. O.G. programmed the codes to analyze the experimental data.

\section*{Conflicts of interest}
There are no conflicts to declare.
\section*{Acknowledgements}
We thank Erick Burgos for helpful discussions.
P.M.O, O.G, D.R and F.G.-L. have received support from the ANID – Millennium Science Initiative Program – NCN19 170, Chile. F.G.-L. was supported by Fondecyt Iniciación No.\ 11220683.
F.G-L. and H.L. thanks the support of DAAD award 91751172.


\balance


\bibliography{Referencias} 
\providecommand*{\mcitethebibliography}{\thebibliography}
\csname @ifundefined\endcsname{endmcitethebibliography}
{\let\endmcitethebibliography\endthebibliography}{}

\end{document}